\documentclass[aps,prl,reprint,superscriptaddress]{revtex4-2}
\pdfoutput=1

\usepackage[colorlinks = true,
            linkcolor = black,
            urlcolor  = blue,
            citecolor = black,
            anchorcolor = black,
            hypertexnames = false]{hyperref}

\usepackage{graphicx,xcolor,amsmath,booktabs}
\usepackage[euler]{textgreek}

\usepackage[activate={true,nocompatibility},final,tracking=true,kerning=true,spacing=true]{microtype}
\begin{document}

\title{Ultrafast element- and depth-resolved magnetization dynamics probed by transverse magneto-optical Kerr effect spectroscopy in the soft x-ray range}

\author{Martin Hennecke}
\email[]{hennecke@mbi-berlin.de}
\author{Daniel Schick}
\email[]{schick@mbi-berlin.de}
\author{Themistoklis Sidiropoulos}
\author{Felix Willems}
\author{Anke Heilmann}
\author{Martin Bock}
\author{Lutz Ehrentraut}
\author{Dieter Engel}
\author{Piet Hessing}
\author{Bastian Pfau}
\affiliation{Max-Born-Institut f{\"u}r Nichtlineare Optik und Kurzzeitspektroskopie, Max-Born-Stra{\ss}e 2A, 12489 Berlin, Germany}
\author{Martin Schmidbauer}\affiliation{Leibniz-Institut f{\"u}r Kristallz{\"u}chtung, Max-Born-Stra{\ss}e 2, 12489 Berlin, Germany}
\author{Andreas Furchner}\affiliation{Helmholtz-Zentrum Berlin f{\"u}r Materialien und Energie, Division Energy and Information, Schwarzschildstra{\ss}e 8, 12489 Berlin, Germany}
\affiliation{Leibniz-Institut f{\"u}r Analytische Wissenschaften -- ISAS -- e.V., Department Berlin, Schwarzschildstra{\ss}e 8, 12489 Berlin, Germany}
\author{Matthias Schnuerer}
\author{Clemens von Korff Schmising}
\affiliation{Max-Born-Institut f{\"u}r Nichtlineare Optik und Kurzzeitspektroskopie, Max-Born-Stra{\ss}e 2A, 12489 Berlin, Germany}
\author{Stefan Eisebitt}
\affiliation{Max-Born-Institut f{\"u}r Nichtlineare Optik und Kurzzeitspektroskopie, Max-Born-Stra{\ss}e 2A, 12489 Berlin, Germany}
\affiliation{Institut f{\"u}r Optik und Atomare Physik, Technische Universit{\"a}t Berlin, Stra{\ss}e des 17. Juni 135, 10623 Berlin, Germany}

\date{\today}

\begin{abstract}
	We report on time- and angle-resolved transverse magneto-optical Kerr effect spectroscopy in the soft x-ray range that, by analysis via polarization-dependent magnetic scattering simulations, allows us to determine the spatio-temporal and element-specific evolution of femtosecond laser-induced spin dynamics in nanostructured magnetic materials.
	In a ferrimagnetic GdFe thin film system, we correlate a reshaping spectrum of the magneto-optical Kerr signal to depth-dependent magnetization dynamics and disentangle contributions due to non-equilibrium electron transport and nanoscale heat diffusion on their intrinsic time scales. 
	Our work provides a quantitative insight into light-driven spin dynamics occurring at buried interfaces of complex magnetic heterostructures, which can be tailored and functionalized for future opto-spintronic devices.
\end{abstract}

\maketitle

The recent advances in ultrafast magnetism research \cite{RevModPhys.82.2731,CARVA2017291} have made it possible to develop the understanding of laser-driven spin dynamics from microscopic processes towards macroscopic functionality in complex systems including charge and spin transport~\cite{Malinowski2008,Battiato2010, Schellekens2014, Igarashi2020, VanHees2020} as well as interactions with spatially extended quasi-particles such as phonons and magnons~\cite{Melnikov2008, Afanasiev2014, Nova2017, Maehrlein2018,Dornes2019,Zhang2020,Windsor2021,Tauchert2022}. 
These findings are exploited in today's opto-spintronics where magnetic order is optically controlled in buried layers and across multiple interfaces of tailored magnetic nanostructures.
The combination of spatio-temporal spin manipulation and structure design on the nanoscale is highly relevant for novel ultrafast and energy-efficient information technology~\cite{Vedmedenko2020}, thermoelectrics~\cite{Bauer2012}, and THz emitters~\cite{Kampfrath2013}.
While common optical and transport-based magnetometry only provides an indirect access to the spatial dependence of the relevant non-local spin dynamics, ultrafast resonant magnetic soft x-ray scattering~\cite{Kortright2013} is a unique technique, which combines magnetic contrast with element selectivity and nanoscale depth resolution, utilizing the short wavelength of the radiation with access to core level resonances.
This has been exploited in first steps towards ultrafast magnetization depth profiling at large scale facilities~\cite{Sant2017, Chardonnet2021}, however, so far without being able to provide a quantitative analysis of the spatially inhomogeneous evolution of the magnetization with sub-picosecond temporal resolution.
One challenge in order to exploit resonant magnetic scattering for nanoscale magnetization profiling is related to the complex interplay of angle- and/or photon energy-dependent absorption, refraction, interlayer reflections, and interference effects.
To that end, it is inevitable to compare the resulting complex scattering patterns with magnetic scattering simulations~\cite{Macke2014, Elzo2012} relying on carefully determined atomic and magnetic form factors, especially for photon energies in close vicinity to core-to-valence-band transitions.
Continuous progress in laser-driven table-top light sources based on high-harmonic generation (HHG) has started to provide access to the relevant soft x-ray regime -- for a long time the exclusive domain of large-scale facilities such as synchrotrons and free-electron lasers.
In particular, HHG with control over the light polarization \cite{Vodungbo2011,Kfir2015,Fan2015, Lambert2015} has enabled laboratory-based experiments on magnetization dynamics with temporal resolutions down to the femto- \cite{Mathias2012,Hofherr2020,Tengdin2020,Willems2020,Zayko2021} or even attosecond regime~\cite{Siegrist2019}.
Above all, the broadband properties of HHG radiation allow measurements of large spectral regions in a single acquisition, providing simultaneous access to characteristic spectral features at their intrinsic time scale.

\begin{figure*}
	\includegraphics[width=1\linewidth]{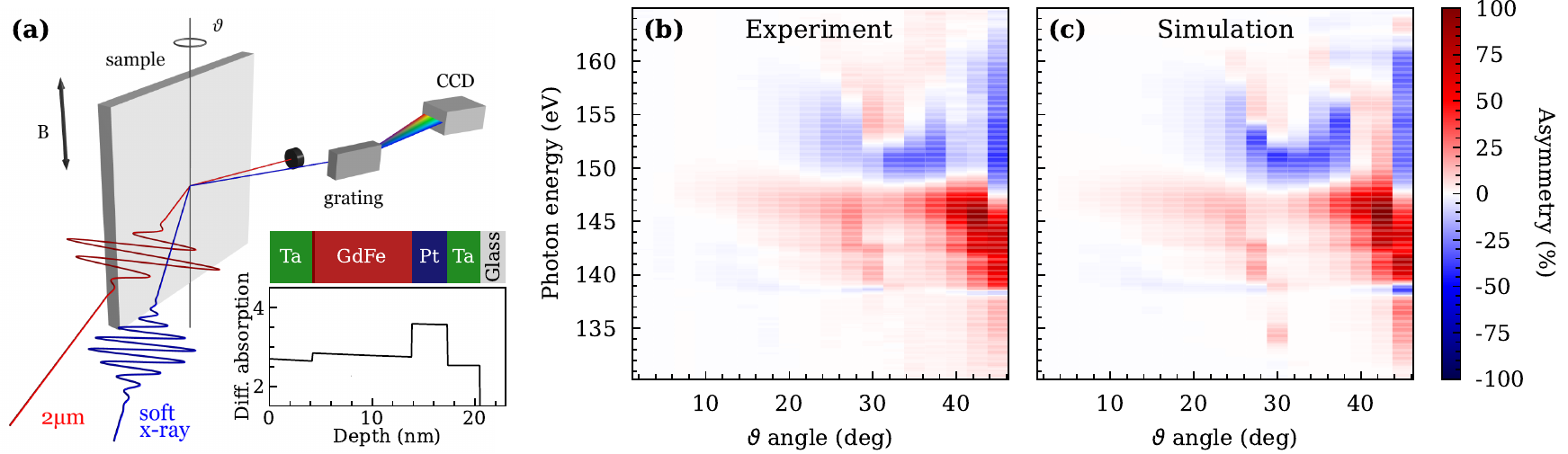}
	\caption{(a) Schematic illustration of the $\vartheta$--$2\vartheta$ spectroscopy setup used for the angle- and time-resolved experiments. 
	The inset shows the depth-dependent differential absorption of the 2.1\,\textmu m pump pulse (incident from the left side onto the Ta layer) that drives the magnetization dynamics.
	(b)~Static angle-resolved TMOKE asymmetry spectra (color scale) of the studied GdFe sample measured by soft x-ray pulses in the photon energy range across the Gd $\text{N}_{5,4}$ resonance. 
	(c) Simulation of the static angle-resolved TMOKE asymmetry spectra fitted to the experimental data in panel (b). 
	\label{1_Figure}}
\end{figure*}

In this work, we study the spatial evolution of the ultrafast magnetization dynamics in a ferrimagnetic Ta/GdFe/Pt nanolayer system; a model system exhibiting intriguing magnetic functionality, such as single shot all-optical switching \cite{Stanciu2007,Radu2011} and self-induced spin-orbit torques \cite{Cespedes-Berrocal2021,Krishnia2021}, both relevant for potential spintronics applications. 
We conduct femtosecond time- and angle-resolved transverse magneto-optical Kerr effect (TMOKE) spectroscopy at the Gd $\text{N}_{5,4}$ resonance around 150\,eV photon energy, selectively probing the magnetization of the Gd sublattice.
Our analysis reveals significant differences in the ultrafast evolution of the TMOKE asymmetry for different photon energies.
By comparison to polarization-dependent magnetic scattering simulations, we can quantitatively relate these spectral changes to transient, depth-dependent de- and remagnetization profiles within the GdFe layer after photoexcitation.
Analysis of the evolving magnetization depth profiles allows us to disentangle femtosecond dynamics dominated by non-equilibrium electron transport ($\leq 100$\,fs) and nanoscale heat diffusion on a picosecond time scale ($\geq 1$\,ps). 
Based on our experimental data, we hence rule out significant contributions due to femtosecond non-local spin transport phenomena, but observe the emergence of a magnetization gradient within GdFe within approximately 1\,ps, induced by heat injection at the interface with the buried seed layer. 
Our results emphasize the importance of a careful analysis of magnetic scattering data, as inhomogeneous spin dynamics in layered magnetic systems results in a complicated spectral dependence of the TMOKE observable. 
In turn, this allows us to disentangle local and non-local processes on ultrafast time scales.
Importantly, our findings directly correlate experimental observables with functionality in nanoscale device structures, e.g., controlled by charge or spin currents as well as nanoscale heat transfer.

The TMOKE measurements are carried out employing a combined $\vartheta$--$2\vartheta$ reflectometry and spectroscopy setup as schematically illustrated in Fig.~\ref{1_Figure}(a). 
The probing soft x-rays are provided by a laboratory HHG-based light source. 
The HHG process is driven by a high average power (29\,W) optical parametric chirped-pulse amplifier system emitting in the mid-infrared (MIR) spectral range at 2.1\,\textmu m center wavelength with 27\,fs full-width at half-maximum (FWHM) pulse duration.
The source delivers $\leq 27$\,fs FWHM soft x-ray pulses at 10\,kHz repetition rate over a broad and continuous spectrum of which we use the 100--200\,eV region in this work~\cite{moerbeck-bock_opcpa_2021}.
A fraction of the MIR beam is separately guided over a variable delay line, providing synchronized pump pulses for inducing ultrafast magnetization dynamics in the sample in time-resolved pump-probe experiments. 
Both pulses are $p$-polarized and focused onto the sample almost collinearly, facilitating a temporal resolution of $\approx 50$\,fs. 
The soft x-ray pulses are incident under a glancing angle $\vartheta$ on the sample and reflected into the spectrometer placed at $2\vartheta$ with respect to the soft x-ray beam axis. 
The spectrum is horizontally dispersed and focused onto a CCD camera upon reflection by a variable line spacing (VLS) grating, achieving a photon energy resolution down to 0.5\,eV (for more experimental details, see Supplemental Material). 

The sample investigated is an amorphous ferrimagnetic $\text{Gd}_{24}\text{Fe}_{76}$ alloy with in-plane magnetic anisotropy deposited by magnetron sputtering on a 400\,\textmu m thick glass substrate, seeded by a Pt layer and capped with Ta to prevent oxidation of the ferrimagnetic layer.  
All measurements were carried out at room temperature ($\approx 300$\,K), which is below the ferrimagnetic compensation temperature of the studied GdFe layer \cite{Ostler2011}.
Static angle-resolved TMOKE spectra of the sample measured in the photon energy range across the Gd~$\text{N}_{5,4}$ resonance for angles of incidence ranging from $\vartheta = 2.5^\circ$ up to $45^\circ$ are shown in Fig.~\ref{1_Figure}(b). 
The magnetic asymmetry is calculated from two spectra recorded for opposite directions of a saturating in-plane magnetic field applied perpendicular to the $p$-polarization axis of the soft x-ray pulses. 
The experimental data is reproduced with high accuracy by the magnetic scattering simulations shown in Fig.~\ref{1_Figure}(c)~\cite{schick_udkm1dsim_2021, Elzo2012}.
The simulations include the photon energy-dependent reflectivity of the $p$-polarized soft x-rays resulting from the element-specific atomic and magnetic form factors, taking into account both the structural properties of the individual layers, i.e., thickness, density, and roughness, as well as the magnetic moments in the GdFe layer.
By numerically calculating the polarization-dependent wave propagation for each angle of incidence and photon energy, the varying absorption, refraction, and probing depth in the vicinity of the atomic resonance is taken into account for both magnetization states of the sample.
Simulating the $\vartheta$--$2\vartheta$ spectroscopy data thus allows us to fit both the structural parameters of the sample as well as the GdFe layer's equilibrium magnetization component perpendicular to the $p$-polarization axis of the probing soft x-ray pulses (see Supplemental Material for more details about the simulations and fits).
The resulting sample composition (thickness in nm) is Ta(4.2)/Gd$_{24}$Fe$_{76}$(9.7)/Pt(3.5)/Ta(3.2)/Glass.
Mapping the magnetization profile within the GdFe layer with 0.2\,nm resolution, the fit converges for a static magnetization profile that slightly decreases towards the interfaces with both the Ta cap and the Pt seed layers.
The high agreement between the experimental data and simulation shown in Fig.~\ref{1_Figure} proves the ability to determine quantitative magnetization depth profiles and structural parameters with sub-nanometer resolution.

\begin{figure}
	\includegraphics[width=1\linewidth]{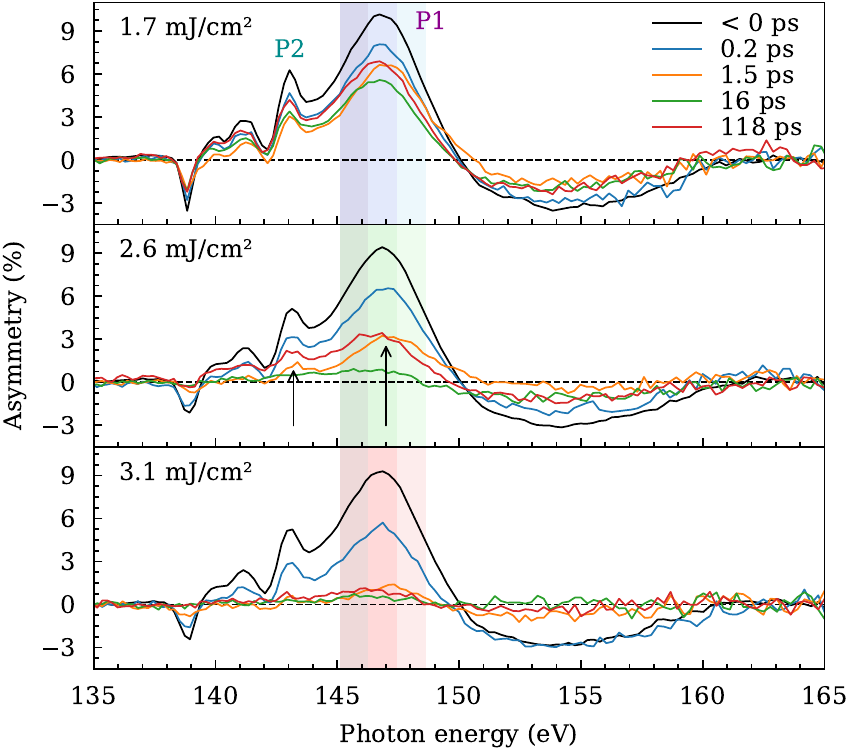}
	\caption{
	Transient TMOKE asymmetry spectra at $\vartheta = 20^\circ$ as a function of incident excitation fluence and pump-probe delay. 
	Each plot shows the equilibrium state (black) as well as a series of transient spectra (colored) recorded at different delays after excitation.
	The shaded areas indicate different photon energy intervals for each fluence (blue, green, red) over which the asymmetry was integrated to obtain the time traces shown in Fig.~\ref{3_Figure}(b).
	The arrows mark the time evolution of the main (P1) and neighboring (P2) asymmetry peaks.
	\label{2_Figure}}
\end{figure}

Time-resolved pump-probe studies were carried out at a soft x-ray probing angle of $\vartheta = 20^\circ$.
The evolution of the asymmetry measured in the photon energy range across the Gd $\text{N}_{5,4}$ resonance is shown in Fig.~\ref{2_Figure} for selected delays and excitation fluences. 
Following the incidence of the pump pulse at 0\,ps, the magnitude of the magnetic asymmetry decreases with respect to the unexcited equilibrium state due to the laser-induced demagnetization of the GdFe layer. 
\begin{figure}
	\includegraphics[width=1\linewidth]{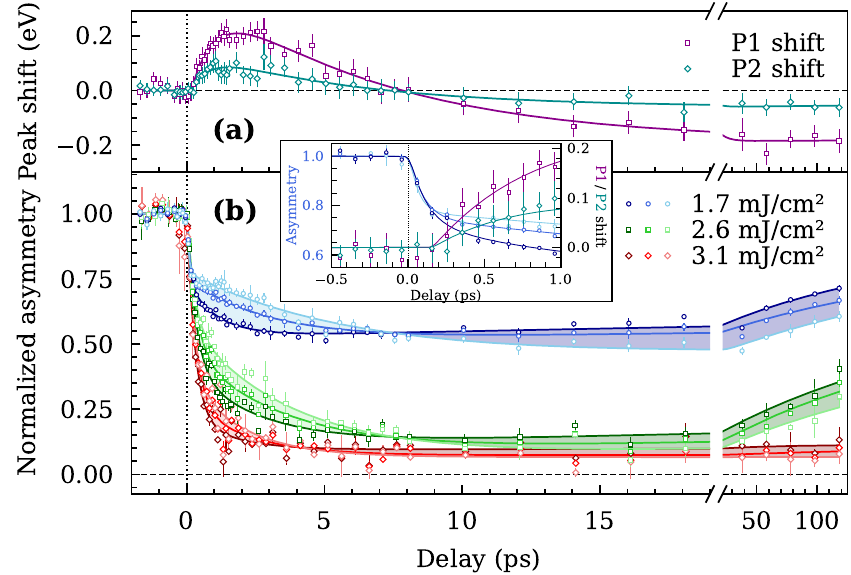}
	\caption{(a)~Transient shift of the main (P1) and its neighboring (P2) asymmetry peak upon 1.7\,mJ/cm$^2$ excitation.
	(b)~Integrated asymmetry as a function of excitation fluence and pump-probe delay. 
	The colored data points for each fluence result from integrating over the accordingly colored photon energy intervals shown in Fig.~\ref{2_Figure}. 
	The inset compares the demagnetization dynamics and peak shifts during the first picosecond after excitation. 
	\label{3_Figure}}
\end{figure}
Comparison of the asymmetry spectra for different time delays after excitation further reveals significant changes of their spectral shape, including a shift of the main peak (P1); first towards higher and later towards lower photon energies with respect to the equilibrium position at $<$ 0\,ps. 
In addition, the relative pump-induced amplitude changes also show a pronounced spectral dependence.
This can be seen particularly well for an incident fluence of 2.6\,mJ/cm$^2$ (center panel of Fig.~\ref{2_Figure}), following the time evolution of the peak amplitudes P1 and P2 at early (1.5\,ps) and late (118\,ps) times.
The observed shift and photon energy dependence is further analyzed in Fig.~\ref{3_Figure}.
Panel~(a) shows the transient relative shifts of the asymmetry peaks P1 and P2 upon 1.7\,mJ/cm$^2$ excitation, obtained from a Gaussian multi-peak fit of the time-resolved spectra. 
Significantly different shifts of the two peaks can be observed which after 1--2\,ps approach their maxima of $\approx 200$\,meV and $\approx 100$\,meV, respectively, stretching the spectrum on the photon energy axis. 
As indicated by the inset, the shifting of peaks starts with a delay of $\approx 200$\,fs after the initial, ultrafast drop of the femtosecond demagnetization. 
At later times, a much slower shift towards the opposite direction leads to a sign change after 8\,ps and a long-lived shifted state which is still present after 120\,ps when the magnetization of the system starts to relax back to its initial state. 
These effects have a strong impact on the TMOKE observable, when the asymmetry signals are integrated within a finite spectral bandwidth around the asymmetry peak of an element-specific resonance, as it is usually done for obtaining magnetization transients as a function of delay time.
This is emphasized in Fig.~\ref{3_Figure}(b), which shows the peak-integrated asymmetry as a function of incident excitation fluence obtained by integrating the time-resolved spectra over slightly shifted photon energy intervals as indicated by the shaded areas in Fig.~\ref{2_Figure}. 
The data is fitted with triple-exponential functions as a guide to the eye.
It becomes obvious that the perceived magnetization dynamics of the GdFe layer is highly dependent on the choice of the integration window, resulting in transients that indicate contradicting magnitude and speed of the pump-induced magnetization change [indicated by the filled areas between the time traces shown in Fig.~\ref{3_Figure}(b)].
Our observation shows that the TMOKE asymmetries measured at different photon energies are not proportional to a uniform magnetization averaged over the whole depth of the magnetic layer.
This indicates that the asymmetries emerge from different probing volumes, as the absorption length strongly varies across the giant N$_{5,4}$ resonance. 
The spectral dependence of the pump-induced change therefore strongly suggests depth-dependent magnetization dynamics of the GdFe layer. 
To that end, we have carefully ruled out any non-magnetic contribution, by finding, within our signal-to-noise level, no transient change of the reflectivity spectra, which measure the laser-induced change of the average 4\textit{f} electronic state occupation (see Supplemental Material). 
This in turn implies that the initial ultrafast drop of the localized Gd 4\textit{f} moment is driven by optical excitation of 5\textit{d}6\textit{s} valence electrons and the very strong intra-atomic exchange coupling between localized 4\textit{f} and itinerant 5\textit{d} moments \cite{Wietstruk2011, Rettig2016}.

\begin{figure}
	\includegraphics[width=1\linewidth]{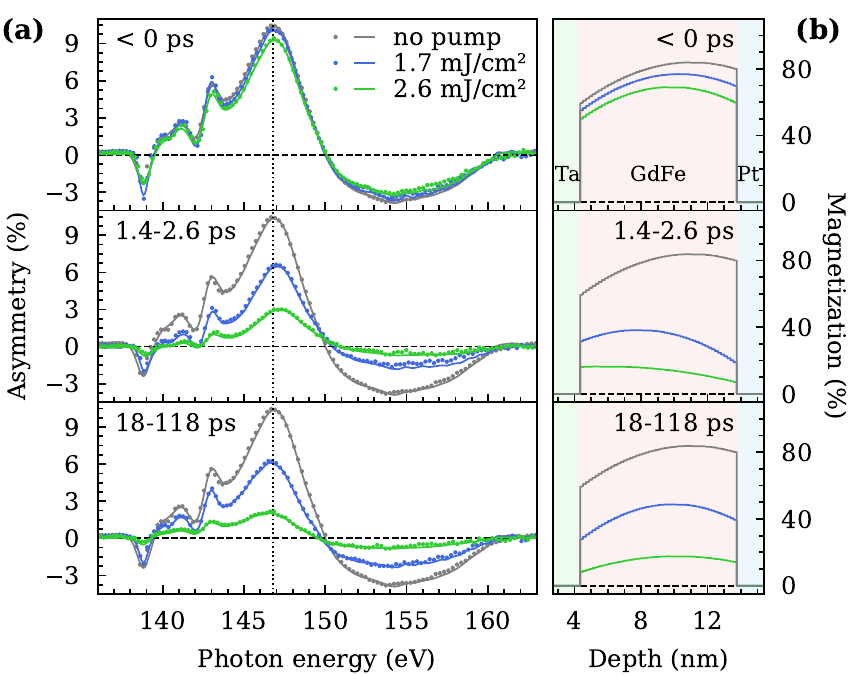}
	\caption{Determination of the magnetization depth profiles from the time-resolved TMOKE spectroscopy measurements. 
	(a)~Recorded transient asymmetry spectra (dots) and fitted simulations (lines) averaged over different time intervals as a function of the incident excitation fluence (black, blue, green).
	The simulation was fitted to the experimental data by varying only the magnetization distribution within the GdFe layer. 
	(b)~Resulting magnetization depth profiles obtained for the respective excitation fluence and time interval. 
	\label{4_Figure}}
\end{figure}

In order to obtain the transient magnetization profiles of the system, the time-resolved experimental data was fitted by simulated asymmetry spectra, varying only the magnetization distribution within the GdFe layer as a fit parameter while keeping the structural parameters fixed that were determined from the previous static angle-resolved measurements.
We model the magnetization profile as a second order polynomial function, as the simplest mathematical description reflecting the asymmetric layer structure of our sample with different cap and seed layers, Ta and Pt, respectively.
Transient changes of the structure, e.g., due to acoustic deformations, result only in marginal changes of the calculated asymmetry spectra for the range of excitation fluences studied and have been neglected, accordingly.
For the sake of clarity and to achieve a better signal-to-noise ratio, the measured asymmetry spectra were averaged over different time intervals, i.e., at early times before the excitation pulse hits the sample and at two later times where the observed shift of the asymmetry peaks is largest in both directions [compare Fig.~\ref{3_Figure}(a)]. 
The simulation is able to reproduce the observed behavior of the transient TMOKE spectra with high accuracy by assuming a unique magnetization depth profile for each time interval, see Fig.~\ref{4_Figure}. 
On the contrary, simulations that assume only a homogeneously distributed magnetization profile or a combination of the latter with transient changes in the atomic, non-magnetic form factors predict significant changes and shifts in the non-magnetic reflectivity spectra (see Supplemental Material), which, as mentioned before, are not observed in the experiment, or otherwise lead to a significantly worse agreement with the experimental data.

The transient evolution of the magnetization profiles can thereby be understood as follows: Before 0\,ps, there is already a slight fluence-dependent decrease of the magnetization with respect to the non-excited equilibrium state due to static heating of the sample.
This is caused by the absorption of the repetitively arriving pump pulses during the pump-probe measurements.
After a pump pulse hits the sample, it causes a thermally-induced demagnetization within the GdFe layer, which is also fluence-dependent. 
The comparison of experiment and simulation shows that, within a time interval of 1.4--2.6\,ps after excitation, the demagnetization is clearly enhanced towards the interface with the Pt seed layer, i.e., on the side of the GdFe layer \textit{not} facing the laser beam.
This observation is in line with a numerical simulation of the depth-dependent absorption profile of the 2.1\,\textmu m pump radiation, predicting the deposited energy density to be largest within the Pt seed layer (see inset of Fig.~\ref{1_Figure} and Supplemental Material for more details). 
In addition to the quantitative determination of magnetization depth profiles, our method also reveals that an inhomogeneous change of the magnetization within the GdFe layer must be accompanied by peak shifts in the asymmetry spectra.
Based on that we carry out a closer inspection of the transient evolution of the evaluated asymmetry amplitude and peak shift.
The decrease of the asymmetry amplitude, as depicted in Fig.~\ref{3_Figure}(b), shows a clear double exponential behaviour with a fast ($\approx 100$\,fs) and a slow ($\approx 1$\,ps) time constant.
These two regimes are commonly referred to as type II dynamics~\cite{Koopmans2010}, and their time scales are determined by the equilibration of the spin system with the electronic and phononic sub-systems.
By comparison to the data in Fig.~\ref{3_Figure} [Panel (a) and inset], we find no sizeable asymmetry peak shift evolving in the fast regime during the first hundreds of femtoseconds.
This indicates that the initial sub-ps demagnetization is spatially \textit{homogeneous} within the GdFe layer and that significant contributions due to non-local inter- or intralayer transport phenomena, such as superdiffusive spin currents~\cite{Battiato2010}, can be excluded.
This conclusion again relies on the notion of very strong intra-atomic exchange interaction between localized and itinerant electrons of Gd, as required to rationalize the observed ultrafast drop of the Gd 4\textit{f} moment within $<$200\,fs.  
Such super-diffusive currents have been shown to lead to significant inhomogeneities in the magnetization profile as well as to interface spin accumulation lasting a few hundreds of femtoseconds after excitation \cite{Battiato2012,Eschenlohr2013,Kampfrath2013}. 
In contrast, the significant peak shifts evolving on time scales of 1--3\,ps go along with the slower time scale of the type II dynamics emerging from electron-phonon thermalization.
This strongly suggests that, due to the layer-dependent laser excitation, phononic heat transfer from the Pt layer into the GdFe layer on a few ps time scale is the main driver for the evolving inhomogeneous magnetization profile. 
In contrast, at later times (18--118\,ps) when the magnetization starts to relax back to the equilibrium state, we observe a negative peak shift caused by a reversed magnetization profile.
This can be rationalized by an opposite thermal gradient evolving due to slow heat dissipation into the glass substrate.
Finally, we note that, beside heat diffusion, thermally driven spin currents on ps time scales could also contribute to depth-dependent magnetization dynamics, caused for example by the spin-dependent Seebeck Effect~\cite{Slachter2010,Choi2014}.

In conclusion, we followed the spatio-temporal and element-selective evolution of magnetization depth profiles in a ferrimagnetic GdFe nanolayer sample by applying ultrafast angle-dependent TMOKE spectroscopy in the soft x-ray range.
The comparison of the experimental data to magnetic scattering simulations provides a direct link between the photon energy-dependent temporal evolution of the TMOKE asymmetry and the spatially inhomogeneous magnetization dynamics.
With a temporal resolution only limited by the laser pulse durations, our experiment is able to resolve depth- and layer-dependent spin transport from picosecond heat transport in nanostructures down to the ultrashort time scales governed by non-thermal, spin-polarized electron currents.  
In general, this allows us to distinguish the relevant local and non-local processes during ultrafast de- and remagnetization based on their intrinsic temporal and spatial fingerprints.

\begin{acknowledgments}
	We thank Ilie Radu for valuable feedback and discussions.
	Funding from the German Research Foundation (DFG, Germany) through CRC/TRR 227 project A02 (project ID 328545488) and from the European Union through EFRE project 1.8/10 and 1.8/15 is gratefully acknowledged. 
	We thank Karsten Hinrichs (Leibniz-Institut f{\"u}r Analytische Wissenschaften -- ISAS -- e.V.) for cooperation via the \textit{Application Lab for Infrared Ellipsometry} (funded by European Union grant EFRE 1.8/13), and Sven Peters (Sentech Instruments) for his expertise in VIS ellipsometry.
\end{acknowledgments}

%
	
\onecolumngrid
\clearpage

\appendix
\renewcommand{\figurename}{\fontsize{11}{22}\selectfont SUPPL. FIG.}
\renewcommand{\tablename}{\fontsize{11}{22}\selectfont SUPPL. TABLE}
\setcounter{equation}{0}
\setcounter{figure}{0}
\setcounter{table}{0}
\fontsize{12}{24}\selectfont

\section{\texorpdfstring{\Large{SUPPLEMENTAL MATERIAL}}{SUPPLEMENTAL MATERIAL}}
\subsection{\texorpdfstring{\large 1. Experimental techniques and data acquisition}{1. Experimental techniques and data acquisition}} \label{section:technique}

The angle- and time-resolved TMOKE measurements were carried out employing a $\vartheta$--$2\vartheta$ spectroscopy setup. 
The soft x-rays for probing the element-specific magnetic moments are provided by a laboratory high-harmonic generation (HHG) based light source operated with noble gas argon ($\approx$ 300\,mbar gas pressure). 
The HHG process is driven by a 2.1\,\textmu m, 27\,fs full-width at half-maximum (FWHM) high average power mid-infrared (MIR) optical parametric chirped pulse amplifier (OPCPA) system and generates $p$-polarized $\leq$ 27\,fs soft x-ray pulses at 10\,kHz repetition rate, covering a broad and continuous spectrum in the range of 100--200\,eV (photon flux $\approx 10^9$\,photons/eV/s @ 150\,eV)~\cite{moerbeck-bock_opcpa_2021_supp}.
The spectroscopy setup consists of a two-circle $\vartheta$--$2\vartheta$ goniometer, allowing both sample ($\vartheta$) and spectrometer ($2\vartheta$) to be placed under angles of 0--$90^\circ$ with respect to the optical axis, covering both transmission and specular reflection geometries for $\vartheta$ angles of 0--$45^\circ$ (measured with respect to the sample plane).
The soft x-ray pulses are focused at the rotation center of the goniometer upon reflection by a toroidal mirror placed at its focal distance of 1000\,mm under a grazing angle of $5^\circ$, leading to a focal spot size of $70\times100$\,\textmu m$^2$ FWHM. 
When carrying out experiments in reflection geometry, the horizontal projection of the soft x-ray spot on the sample depends on the $\vartheta$ angle of incidence and scales with $1/\sin \vartheta$, leading to a spatial footprint of, e.g., $205\times100$\,\textmu m$^2$ FWHM for $\vartheta=20^\circ$. 
The $2\vartheta$ spectrometer consists of a variable line spacing (VLS) grating with central line density of 1200\,l/mm (Hitachi 001-0437) that is placed at its focal length (235\,mm) under a grazing angle of $3^\circ$ with respect to the optical axis of the transmitted or reflected soft x-ray pulses. 
The first and higher order reflection of the VLS grating horizontally disperses and focuses the soft x-ray spectrum on an in-vacuum CCD camera (Greateyes GE 1024 256 BI), resulting in a photon energy resolution down to 0.5\,eV. 
In order to normalize the measured soft x-ray spectra to minimize the influence of intensity fluctuations of the HHG process, 10\% of the soft x-rays are extracted after the toroidal mirror by a beam-splitting membrane (50\,nm thick polyvinyl butyral, PVB) and guided over a reference spectrometer, mirroring the aforementioned components. 
The detectors are protected from parasitic stray light by covering the CCDs with a 200\,nm thin silver filter.

For carrying out time-resolved pump-probe experiments, a fraction ($\approx$ 10\%) of the $p$-polarized MIR beam that drives the HHG process is coupled out before the HHG cell and separately guided over a variable delay line, serving as an intrinsically synchronized pump beam for exciting the sample.
The pump pulses are focused to the sample plane via a refractive optical lens, resulting in a focal spot size of $570\times490$\,\textmu m$^2$ FWHM. 
The MIR pump and soft x-ray probe pulses hit the sample almost collinearly, with the angle of incidence of the pump being $8^\circ$ more grazing compared to the probe. 
In case of a probing angle of $\vartheta=20^\circ$, this leads to a projection of $2740\times490$\,\textmu m$^2$ FWHM parallel to the sample plane, clearly overfilling the probed area which thus can be considered as homogeneously pumped. 
The spatial overlap between pump and probe is achieved by guiding the MIR pulses through a pinhole placed at the focus of the soft x-ray probe and further improved directly on the sample by scanning the pump spot position with an overlap mirror, optimizing for the largest pump-induced change.
The magnetic TMOKE contrast is achieved by alternating a saturating in-plane magnetic field applied to the sample between both directions perpendicular to the $p$-polarization axis of the probing soft x-ray pulses, which also restores the initial magnetization state of the system after each pump-probe cycle.

\begin{figure*}
	\includegraphics[width=1\linewidth]{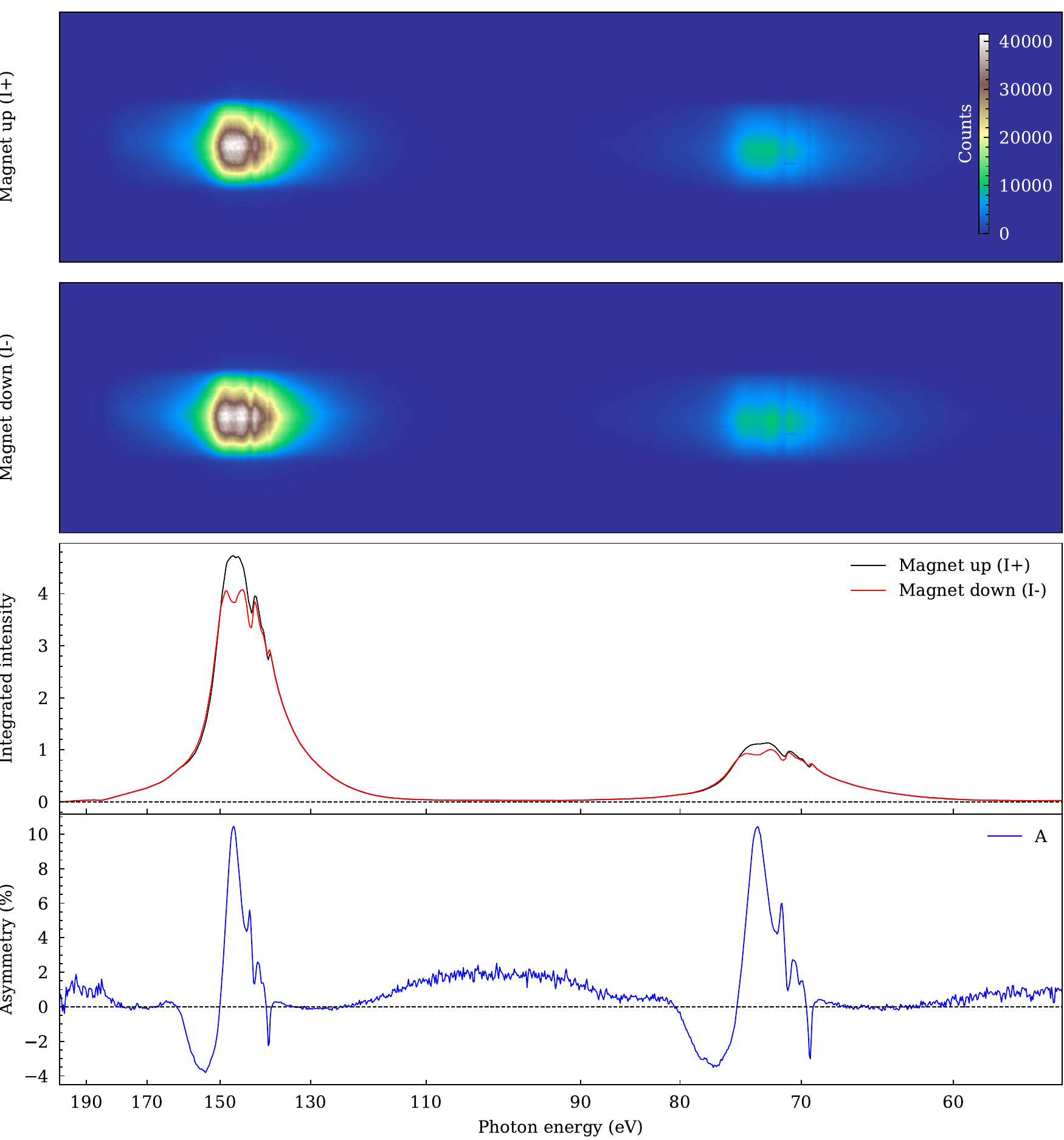}
	\caption{CCD images $I_+$ and $I_-$ of the reflected photon-energy dispersed soft x-ray spectra of the studied GdFe sample recorded for opposite up and down magnetic field directions, respectively, under an angle of incidence $\vartheta=20^\circ$.
	The spectra contain both the first and second order reflection from the VLS grating.
	The lower two panels show the vertically integrated CCD images and the magnetic asymmetry.
	\label{Suppl_1_Figure}}
\end{figure*}

Suppl.~Fig.~\ref{Suppl_1_Figure} shows the CCD images $I_+$ and $I_-$ of the reflected photon-energy dispersed soft x-ray spectra of the studied GdFe sample recorded for opposite up and down magnetic field directions, respectively, under an angle of incidence $\vartheta=20^\circ$, averaged over 21 acquisitions with an illumination time of 20\,s each. 
The duplicate appearance of each spectrum in the CCD images is due to the first and second order reflection of the VLS grating. 
The line plots correspond to the vertical integration of the spectra, from which the magnetic TMOKE asymmetry $A$ is then calculated as $A=(I_+-I_-)/(I_++I_-)$.
The photon energy axis of the recorded images is calculated from the first order reflection of the line grating, depending on its grazing incidence angle and the exact camera distance.
Since we are interested in the asymmetry spectra $A$, any variation of the spectral density per CCD pixel cancels out.
We assign a single wavelength $\lambda$ to each pixel of the CCD according to the reflection grating formula:
\begin{equation} \label{eq:reflgrating}
m \lambda=\frac{1}{g} \left(\sin{\vartheta_{\text{in}}}+\sin{\vartheta_{\text{out}}}\right)
\end{equation}
Here, $m$ denotes the order of reflection, $g$ the line density of the grating, $\vartheta_{\text{in}}$ the grating angle of incidence and $\vartheta_{\text{out}}$ the angle under which the reflection is observed. 
For the dispersed first order reflection, $\vartheta_{\text{out}}$ can be expressed by the camera distance $r$ and the position $\Delta x$ of each pixel ($26\times26$\,\textmu m$^2$) on the CCD relative to the specular (zeroth order) reflection $x_0$:
\begin{equation} \label{eq:firstordertheta}
\vartheta_{\text{out}} = \arctan{\frac{r}{\Delta x + x_0}}
\end{equation}
As for specular reflection $\vartheta_{\text{in}}=\vartheta_{\text{out}}$, $x_0$ can be substituted by $x_0 = r/\tan{\vartheta_{\text{in}}}$. 
Combining Eqs.~(\ref{eq:reflgrating}) and (\ref{eq:firstordertheta}) allows to calculate the wavelength from using only known or determinable quantities:
\begin{equation} \label{eq:lambdaformula}
\lambda = \frac{1}{g} \left[\sin{\left(\arctan{\frac{1}{\frac{1}{\tan{\vartheta_{\text{in}}}}+\frac{\Delta x}{r}}}\right)}+\sin{\vartheta_{\text{in}}}\right]
\end{equation}
Small uncertainties in the experimental realization and determination of geometrical parameters that could lead to an offset or distorted photon energy axis were corrected by calibrating the soft x-ray absorption spectroscopy data obtained from a GdFe membrane sample in transmission geometry to a reference measurement carried out with high energy accuracy at the UE112-PGM1 beamline at the synchrotron BESSY II \cite{schiwietz_ue112_pgm-1_2015}, as well as taking the second order reflection of the grating into account, which needs to perfectly match the first order spectrum when the photon energy axis is multiplied by a factor of two.

\begin{figure*}
	\includegraphics[width=0.65\linewidth]{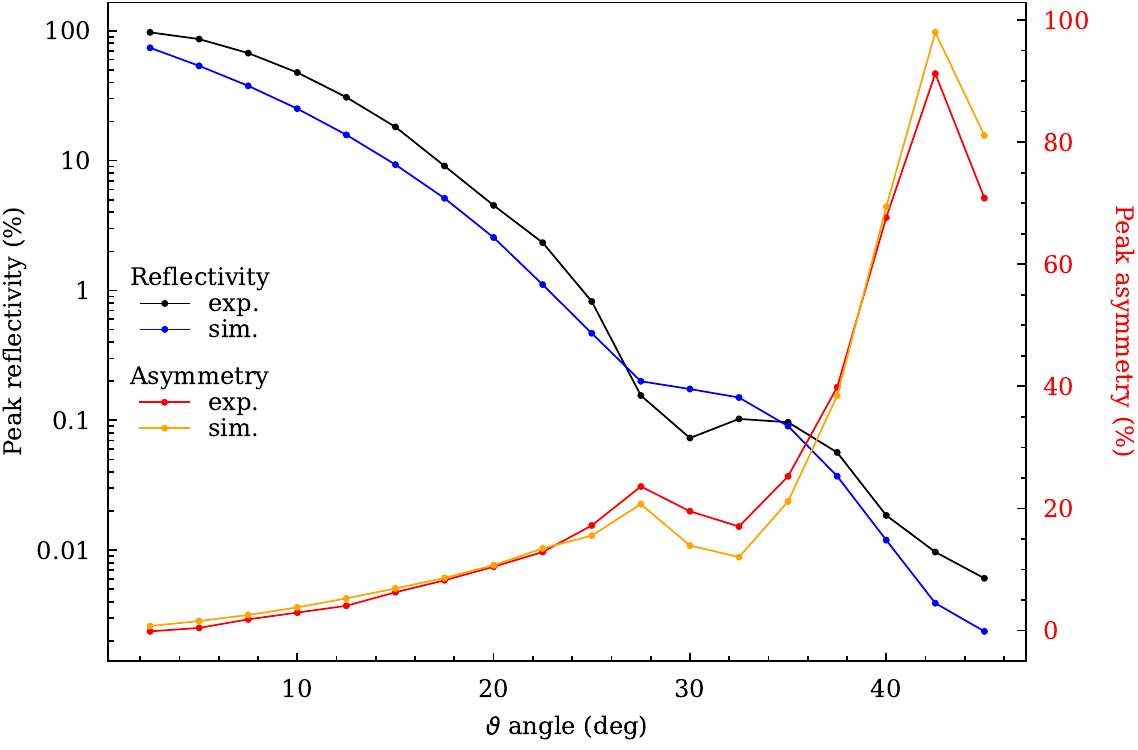}
	\caption{Peak reflectivity (left axis) and TMOKE asymmetry (right axis) of the sample as a function of $\vartheta$ angle, comparing experimental data and simulations (see Section~2).
	\label{Suppl_2_Figure}}
\end{figure*}

The static angle-resolved TMOKE spectroscopy as shown in Fig.~1(b) of the main article was carried out for angles of incidence ranging from $\vartheta = 2.5^\circ$ up to $45^\circ$ in $2.5^\circ$ steps. 
To achieve a reasonable signal-to-noise ratio, the acquisition times had to be adjusted accordingly in order to compensate for the exponentially decaying reflectivity towards higher angles (see Suppl.~Fig.~\ref{Suppl_2_Figure}). 
The number of images recorded for each $\vartheta$ angle and the corresponding illumination times are shown in Suppl.~Table~\ref{tab:acqtimes}. 
In the time-resolved experiments carried out at $\vartheta=20^\circ$, the acquisition times were 30\,s for each of the two images ($I_+$ and $I_-$) which were recorded for each pump-probe delay. 
The transient asymmetry spectra shown in Fig.~2 of the main article were averaged over three consecutive pump-probe delay scans, leading to a total measurement time of 156\,minutes for each excitation fluence.

\begin{table}
	\fontsize{11}{22}\selectfont
	\centering
	\begin{tabular}{cc@{\hspace{10\tabcolsep}}cc@{\hspace{10\tabcolsep}}cc}
		\toprule 
		$\vartheta$($^\circ$) & $N \text{ x } t_{\text{acq}}$ & $\vartheta$($^\circ$) & $N \text{ x } t_{\text{acq}}$ & $\vartheta$($^\circ$) & $N \text{ x } t_{\text{acq}}$ \\ 
		\midrule
		2.5 & $31\times1.5$\,s & 17.5 & $21\times10$\,s & 32.5 & $3\times240$\,s \\
		5.0 & $31\times1.5$\,s & 20.0 & $21\times20$\,s & 35.0 & $3\times300$\,s \\
		7.5 & $31\times2.0$\,s & 22.5 & $16\times30$\,s & 37.5 & $3\times300$\,s \\
		10.0 & $31\times3.0$\,s & 25.0 & $11\times30$\,s & 40.0 & $6\times300$\,s \\
		12.5 & $21\times5.0$\,s & 27.5 & $6\times120$\,s & 42.5 & $6\times300$\,s \\
		15.0 & $21\times7.5$\,s & 30.0 & $6\times120$\,s & 45.0 & $3\times300$\,s \\
		\bottomrule
	\end{tabular} 
	\caption{Number of images $N$ recorded with an illumination time $t_{\text{acq}}$ for each angle $\vartheta$. \label{tab:acqtimes}}
\end{table}

\subsection{\texorpdfstring{\large 2. Static $\vartheta$--$2\vartheta$ spectroscopy and simulations}{2. Static spectroscopy and simulations}}

\begin{figure*}
	\includegraphics[width=0.5\linewidth]{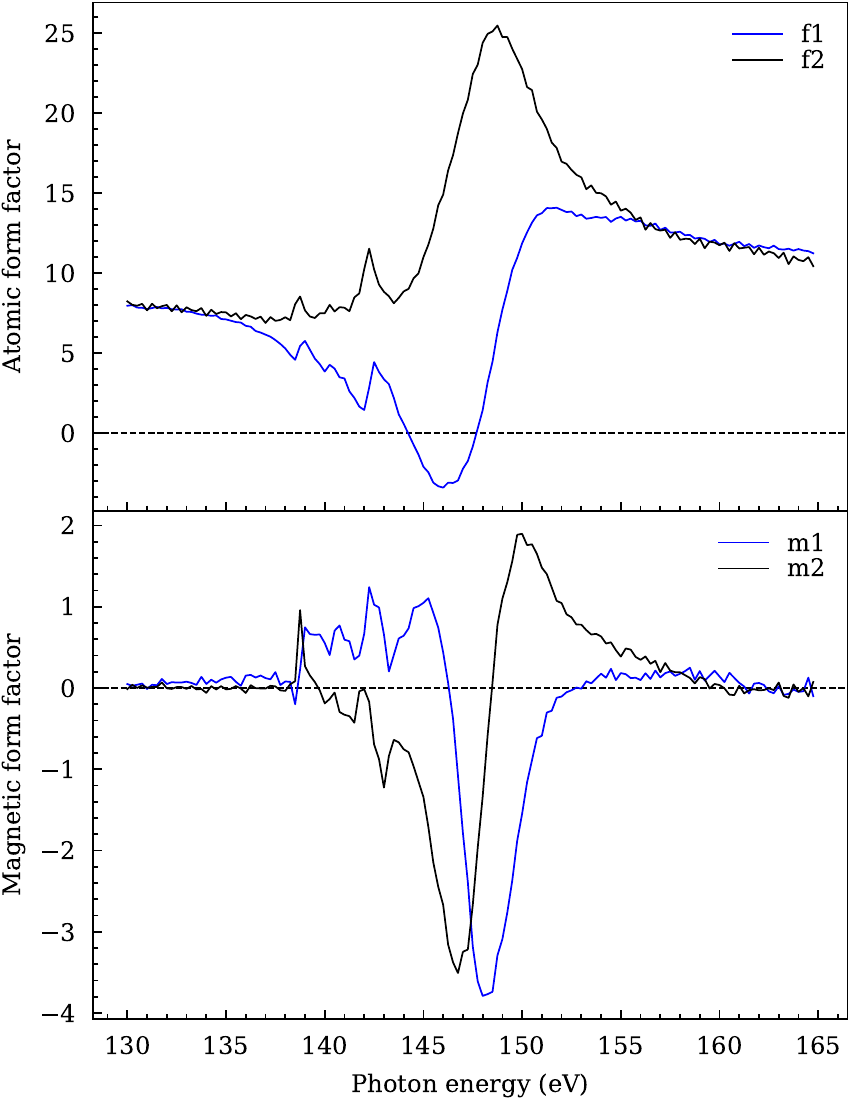}
	\caption{Atomic ($f_1$, $f_2$) and magnetic ($m_1$, $m_2$) form factors of the GdFe layer at the Gd $\text{N}_{5,4}$ resonance.
	\label{Suppl_3_Figure}}
\end{figure*}

The magnetic TMOKE simulations that were used to reproduce the experimental data in Fig.~1 of the main article were carried out using the \textsc{udkm1Dsim} toolbox \cite{schick_udkm1dsim_2021_supp,Elzo2012_supp}, calculating the photon energy-dependent reflectivity of the $p$-polarized soft x-rays from the element-specific atomic and magnetic form factors of the individual layers in the sample. 
Both real and imaginary part of the atomic and magnetic form factors of the GdFe alloy which were used as input for the simulations are shown in Suppl.~Fig.~\ref{Suppl_3_Figure}. 
They were precisely determined from the same reference measurement that was also used for the energy axis calibration of the HHG spectrum, employing an interferometric method that allows to simultaneously retrieve amplitude and phase contrast based on the interference of circularly polarized soft x-rays \cite{hessing_phd_2021}.
The form factors of the glass substrate and surrounding Ta and Pt layers far from their atomic resonances were obtained from the tabulated values of Ref.~\cite{henke_db_1993}.

\begin{figure*}
	\includegraphics[width=0.5\linewidth]{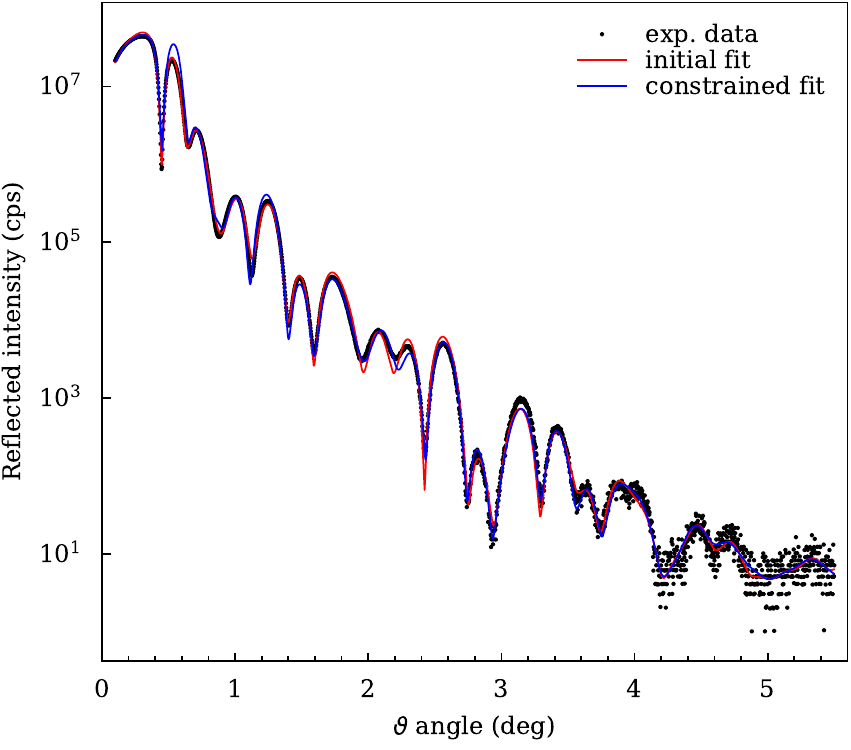}
	\caption{Hard x-ray specular reflectometry (XSR) measurements of the studied GdFe sample, fitted using the software RCRefSimW \cite{zaumseil_rcrefsimw} in order to retrieve the structural properties (thickness, density, roughness) of the individual layers within the sample. 
	Shown is the initial fit with all free parameters (red) in comparison to a constrained fit (blue) closer to the values obtained from fitting the TMOKE data.
		\label{Suppl_4_Figure}}
\end{figure*}

\begin{table}
	\fontsize{11}{22}\selectfont
	\centering
	\begin{tabular}{l@{\hspace{10\tabcolsep}}ccc@{\hspace{10\tabcolsep}}cc@{\hspace{10\tabcolsep}}ccc@{\hspace{10\tabcolsep}}c}
		\toprule
				& \multicolumn{3}{c@{\hspace{10\tabcolsep}}}{XSR (init.)}	& \multicolumn{2}{c@{\hspace{10\tabcolsep}}}{TMOKE} & \multicolumn{3}{c@{\hspace{10\tabcolsep}}}{XSR (constr.)} & \\ 
		Layer	& $t$(nm)	& $\rho$	& $r$(nm)   & $t$(nm)	& $\rho$	& $t$(nm)	& $\rho$    & $r$(nm)   & $\rho_{\text{nom}}$(kg/$\text{m}^3$)	\\ 
		\midrule
		Ta		& 4.04	    & 39\%		    & 0.3       & 4.18	    & 65\%	& 4.51	    & 61\%      & 0.8       & 16,650	\\
		$\text{Gd}_{24}\text{Fe}_{76}$ & 9.93 & 109\%	& 0.3       & 9.66	& 107\%	& 9.62	    & 111\%	    & 0.3       & 7,880\\
		Pt		& 4.20	    & 82\%		    & 0.4       & 3.49	    & 82\%		& 3.02	    & 102\%	    & 0.9       & 21,400	\\
		Ta		& 2.01	    & 88\%		    & 1.1       & 3.16	    & 78\%		& 3.16	    & 85\%	    & 0.4       & 16,650	\\
		Glass 	& -		    & 100\%		    & 0.6       & -		    & 100\%		& -		    & 100\%	    & 0.4       & 2,200	\\
		\bottomrule
	\end{tabular}  
	\caption{Structural parameters of the individual layers in the studied GdFe sample (thickness $t$, percentage of nominal density $\rho$, rms roughness $r$), obtained from angle-resolved XSR measurements and magnetic TMOKE spectroscopy. 
	The values of the nominal densities $\rho_{\text{nom}}$ are obtained from Ref.~\cite{henke_db_1993}.
	For the initial XSR fits, all parameters were free. 
	For the constrained XSR fit, the Ta buffer layer thickness was fixed to the value obtained from the TMOKE data and the Ta densities were constrained to an interval of 60--70\% (cap) and 70--90\% (buffer), respectively. \label{tab:structure}}
\end{table}

\begin{figure*}
	\includegraphics[width=1\linewidth]{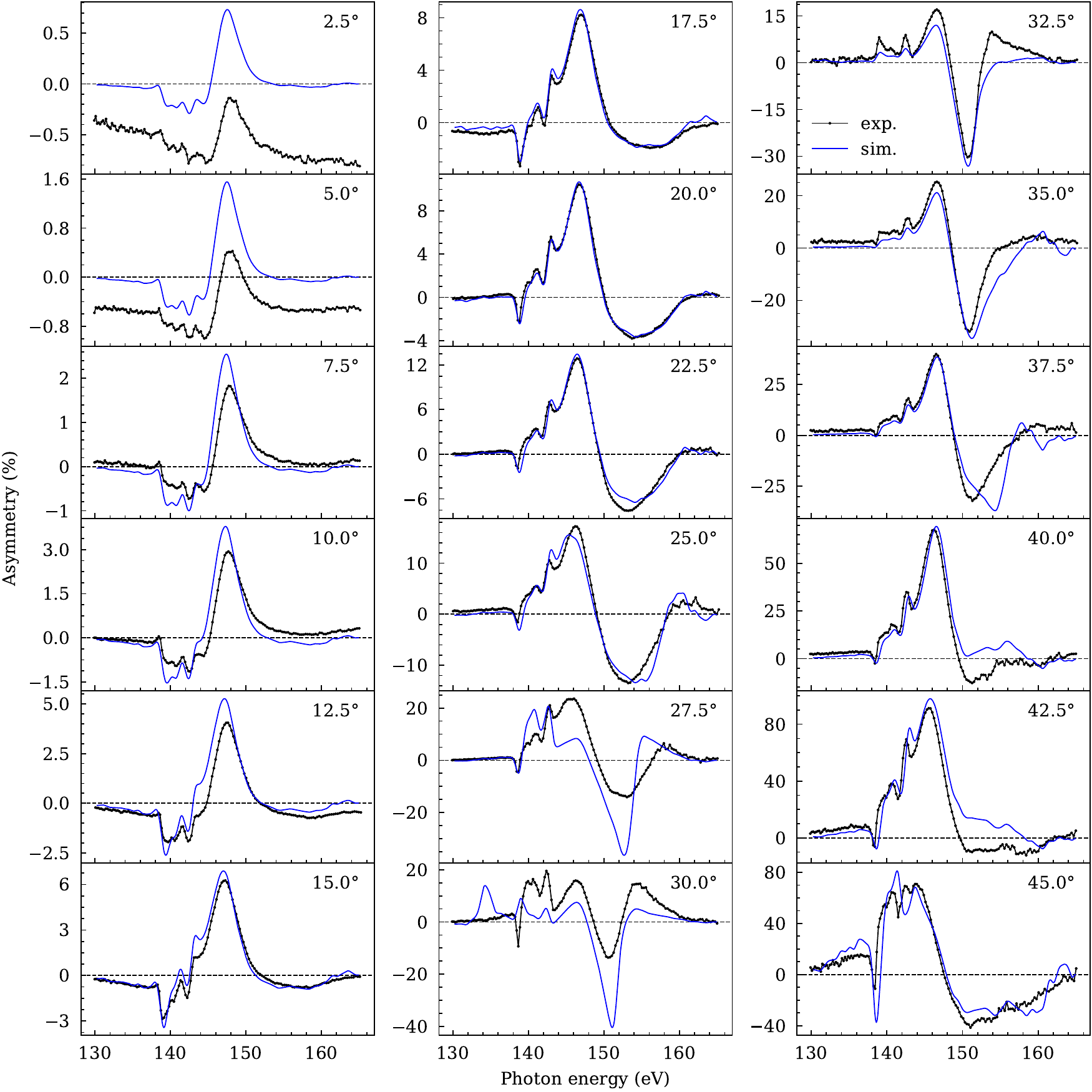}
	\caption{Line plots of the static angle-resolved TMOKE asymmetry of the GdFe sample which is shown as a 2D map in Fig.~1b of the main article. 
	Plotted are the experimentally obtained magnetic asymmetry spectra (black) as a function of $\vartheta$ angle of incidence in comparison to the output of the simulations after fitting the structural parameters of the sample.
	\label{Suppl_5_Figure}}
\end{figure*}

For an accurate simulation of the TMOKE asymmetry and its angle dependence, the exact structural properties of the sample, i.e., thicknesses, densities, and roughnesses of the individual layers in the stack, need to be known. 
In order to determine an initial parameter set for the non-magnetic structural layer properties of the studied sample, hard x-ray specular reflectometry (XSR) was performed on a commercial high-resolution x-ray diffractometer (Rigaku SmartLab 9\,kW \cite{rigaku_smartlab}). Behind a parabolic multilayer mirror, an asymmetric two-fold Ge(220) channel-cut crystal selects the Cu $K\alpha_1$ line (8.047\,keV) and simultaneously collimates the primary beam in the sample reflection plane to about $\Delta\vartheta = 0.008^\circ$. Subsequently, the useful beam is narrowed to 0.1\,mm. The specularly reflected x-ray beam from the sample is detected by a single photon counting area detector (HyPix-3000) using a virtual slit of 0.1\,mm. The experimentally determined x-ray reflectivity was fitted within the framework of dynamical scattering theory applying Parratt’s recursive formalism~\cite{Parratt_1954} (see Suppl.~Fig.~\ref{Suppl_4_Figure}). For that purpose the software package RCRefSimW \cite{zaumseil_rcrefsimw} was used.

Due to a higher sensitivity of the combined angle- and photon energy-resolved TMOKE data in the soft x-ray range for the structure of the very thin layer system, the initial parameters obtained from XSR were further refined by fitting the angle-resolved TMOKE asymmetry spectra with the magnetic scattering simulations (\textsc{udkm1Dsim}). 
This also allows to simultaneously fit the equilibrium magnitude and distribution of magnetic moment in the GdFe layer, which was allowed to be inhomogeneous, modeled by a second order polynomial distribution along the sample normal and allowing magnetically dead layers at the interfaces with the capping and seed layers.
The fit thereby improves for the inclusion of a very thin magnetically dead layer (0.24\,nm) in GdFe at the Ta interface, which is in line with SQUID magnetometry studies of comparable Fe-based alloys capped with Ta~\cite{kowalewski_Ta_2000}.
To limit the computational complexity, the angle-resolved fit of the asymmetry data was carried out only for a subset of eight $\vartheta$ angles (10, 20, 25, 35, 37.5, 40, 42.5 and $45^\circ$). 
While the XSR fits could be improved by allowing parts of the Ta cap layer to be oxidized ($\text{Ta}_2\text{O}_5)$, the TMOKE fit was inconclusive in this regard. 
For simplicity, the magnetic simulations were thus carried out assuming a pure Ta cap layer. 
A comparison of the refined layer thicknesses and densities to the values initially obtained from the XSR measurements is shown in Suppl.~Table~\ref{tab:structure}, showing the largest differences in the Ta buffer layer thickness and cap layer density. 
Please note that the layer roughnesses could only be determined from the XSR measurements, as the magnetic scattering simulations showed only negligible impact on the TMOKE asymmetry.
To assess the significance of the deviations, additional fits of the XSR data were carried out by constraining the most deviating Ta parameters close to the values obtained from the TMOKE data. 
The resulting fit (shown also in Suppl.~Fig.~\ref{Suppl_4_Figure}) still shows good agreement to the experimental data. 
It can thus be concluded that the structure determined by the different techniques agrees within an experimental error of $\leq$~5\,\AA.
Line plots of the magnetic simulations carried out for the determined structure, which are shown as a 2D map in Fig.~1 of the main article, are shown in Suppl.~Fig.~\ref{Suppl_5_Figure}. 
The remaining deviations between the measured and simulated asymmetries can be explained by the footprint of the soft x-ray pulse on the sample scaling with $1/\sin \vartheta$, thus changing the probed area for each angle. 
As a consequence, lateral inhomogeneities of the samples lead to a slight inconsistency that cannot be fully accommodated by assuming a single structure for the whole $\vartheta$ range. 
As the demagnetization dynamics was recorded at $\vartheta=20^\circ$, a higher weight was put on fitting the magnetic TMOKE asymmetry measured at this angle, most accurately describing the structure and magnetization distribution within the volume probed in the time-resolved experiments.

\subsection{\texorpdfstring{\large 3. Transient magnetization depth profiles}{3. Transient magnetization depth profiles}}

\begin{figure*}
	\includegraphics[width=1\linewidth]{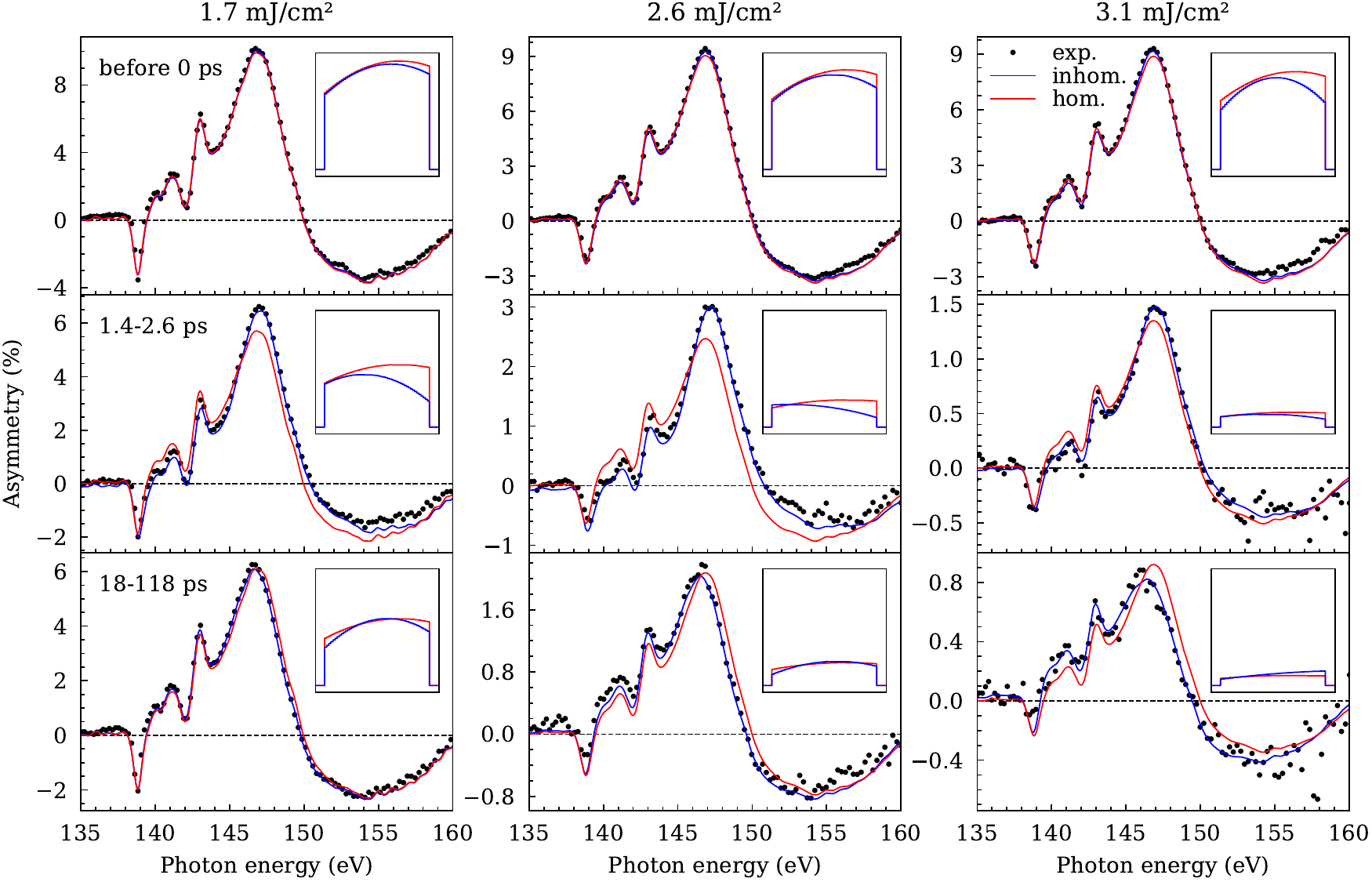}
	\caption{Simulations of the magnetic TMOKE asymmetry fitted to the experimental data recorded for 1.7, 2.6 and 3.1\,mJ/cm$^2$ excitation by either allowing an inhomogeneous change of the magnetization depth profiles (blue curves) or by constraining the fit to a homogeneous demagnetization, scaling only the equilibrium magnetization profile with a constant factor (red curves). 
	The insets show the corresponding magnetization depth profiles. 
	\label{Suppl_6_Figure}}
\end{figure*}

The time-resolved experimental data was fitted by the simulations varying only the magnetization distribution within the GdFe layer and keeping the structural parameters fixed to the values as obtained from the static data. 
As explained in the main article, the transient asymmetry spectra can only be reproduced by assuming a unique magnetization depth profile for each time interval which leads to the conclusion of an inhomogeneous demagnetization of the GdFe layer which is enhanced towards the Pt interface.
Suppl.~Fig.~\ref{Suppl_6_Figure} illustrates the significance of an inhomogenous magnetization change on the simulated asymmetry spectra. 
For clarity, the simulations were also carried out assuming a homogeneous demagnetization by fitting only a constant factor that scales the equilibrium magnetization profile. 
As can be seen for each excitation fluence, this results in a significant deviation from the experimental data, being unable to reproduce the transient spectral shifts and photon energy-dependent changes. 
This also confirms what can already be expected from the peak shifts and deviations between different integration intervals shown in Fig.~3 of the main article, as these would not have been observed when the demagnetization could be described by a homogeneous decay. 
Please note that the magnetization depth profiles evaluated for the highest fluence (3.1\,mJ/cm$^2$) were not shown in the main article, because the strong excitation leads to almost full demagnetization of the GdFe layer over its whole thickness.

\begin{figure*}
	\includegraphics[width=1\linewidth]{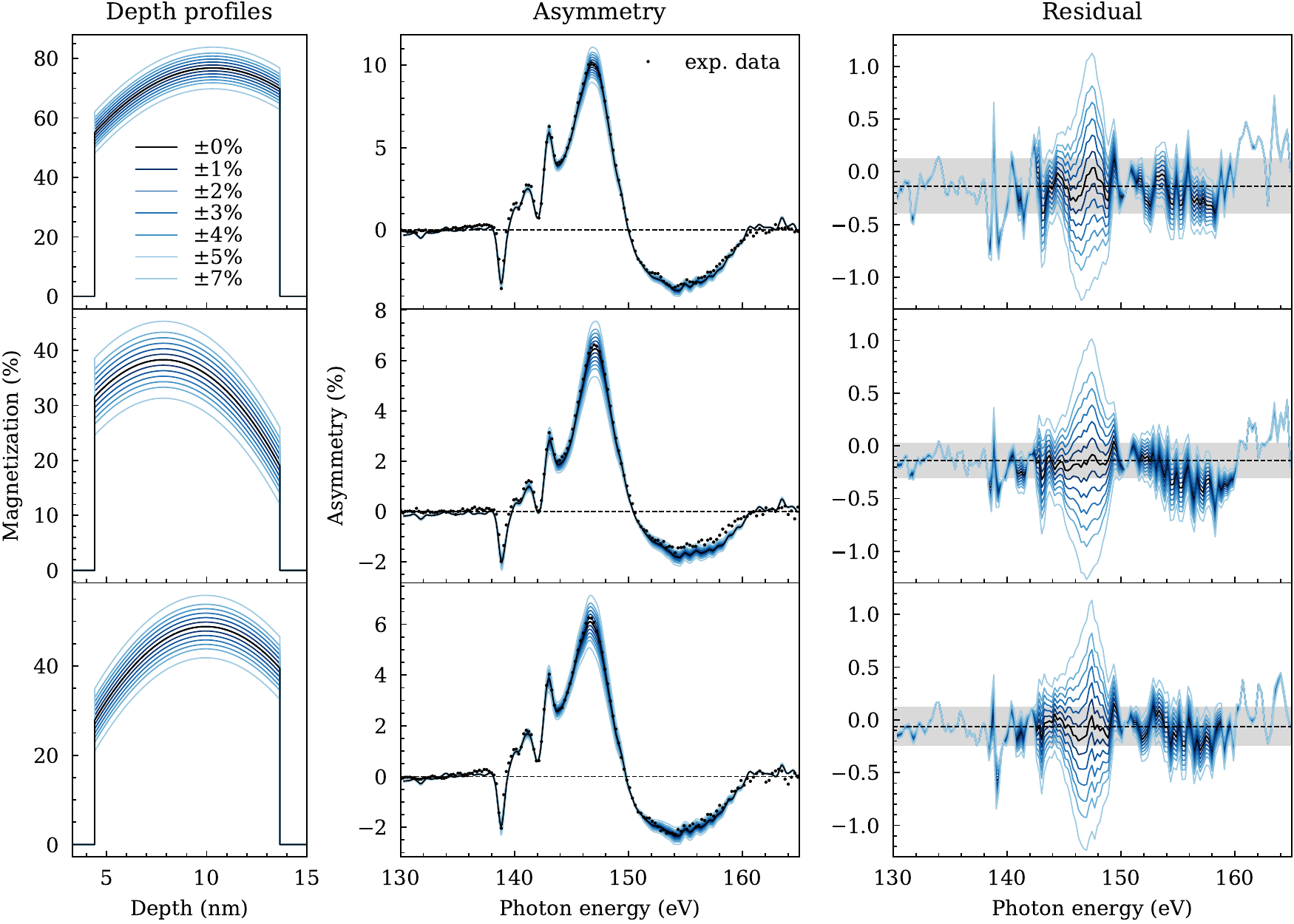}
	\caption{Simulating the impact of adding artificial offsets ($\pm1$--7\,\%) to the best fits (denoted as $\pm0$\,\%) of the magnetization depth profiles on the TMOKE asymmetry, analyzing the corresponding residuals between simulated asymmetry and experimental data. 
	The shaded areas correspond to a 95\,\% confidence interval ($2\sigma$ of the baseline determined in the range of 130--137\,eV). 
	The analysis is carried out for the three different time intervals (upper, mid and lower panels) as discussed in the main article.
	\label{Suppl_8_Figure}}
\end{figure*}

\begin{figure*}
	\includegraphics[width=1\linewidth]{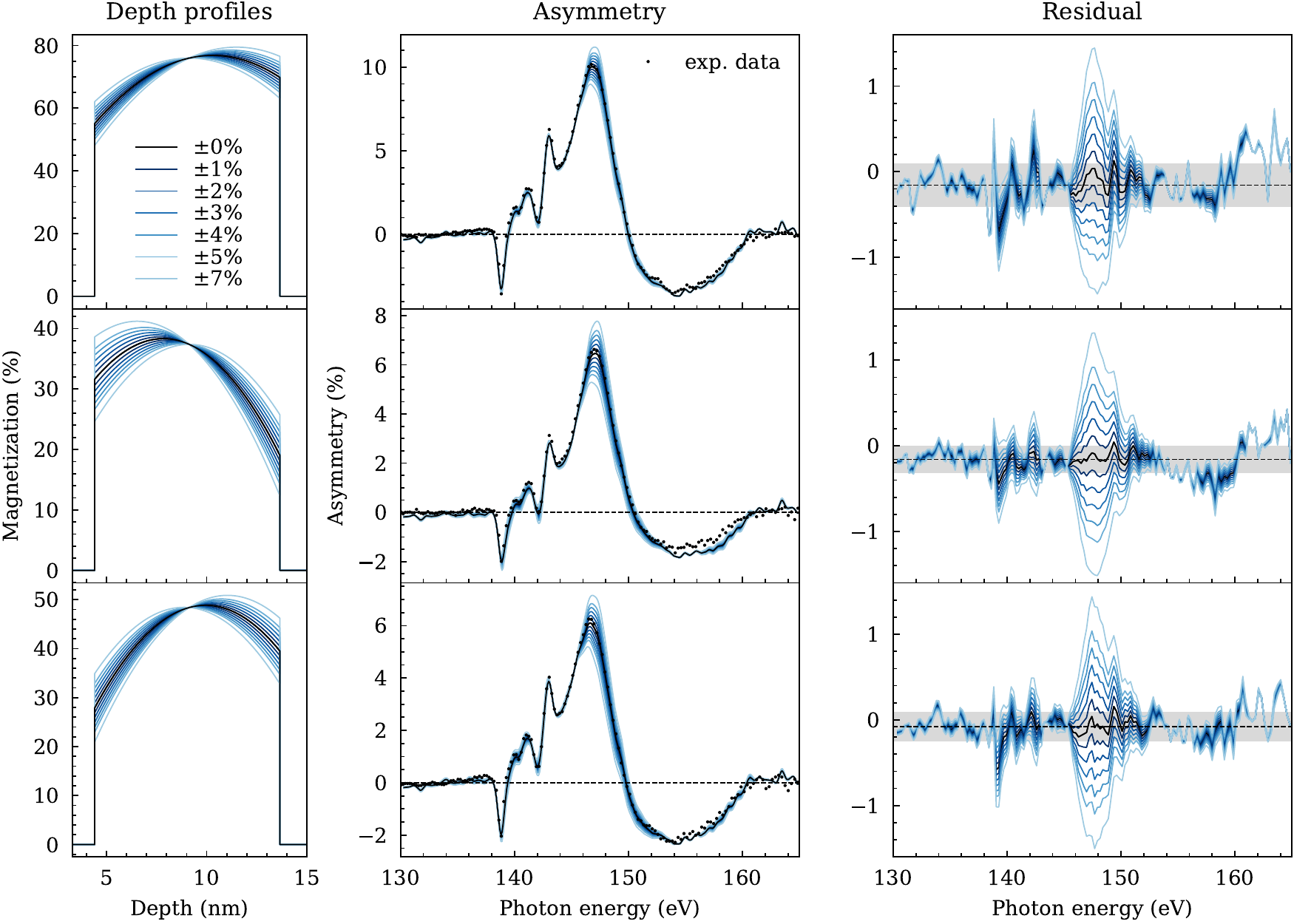}
	\caption{See the caption of Suppl.~Fig.~\ref{Suppl_8_Figure}. 
	Here, the same analysis is carried out for adding artificial slopes instead of offsets to the magnetization depth profiles.
	\label{Suppl_9_Figure}}
\end{figure*}

The uncertainty in the quantitative determination of the magnetization depth profiles is analyzed in Suppl.~Fig.~\ref{Suppl_8_Figure} and \ref{Suppl_9_Figure} by adding artificial offsets and slopes within a range of $\pm1$--7\,\% to the best matching fit of the depth-dependent magnetization distribution. 
The analysis is carried out for all three time intervals that are discussed in the main article and show the impact of the artificially distorted depth profiles on the simulated TMOKE asymmetry as well as the corresponding residuals between simulation and experimental data. 
The latter serve as a metric for the sensitivity of our method and show that for a deviation of $\geq 2$\,\% in magnetization, the residuals along the giant resonance peak ($\approx 148$\,eV) become significantly larger than the 95\,\% confidence interval indicated by the shaded area.
Suppl.~Fig.~\ref{Suppl_9_Figure} further demonstrates the link between a changing slope of the magnetization profile (as happening also during the inhomogeneous demagnetization) and significant shifts observed in the asymmetry spectrum (clearly visible at the right edge of the giant resonance at $\approx 148$\,eV), underlining the high sensitivity of our method differentiating between slope and offset variations.
We therefore estimate the sensitivity of our experiment to be around $\approx 2$\,\% in magnetization. 
However, as the limiting factor is the signal-to-noise ratio of the asymmetry measurement, this could be improved further by increasing the integration times.

In order to assess the impact of choosing a second order polynomial to model the inhomogeneous magnetization profiles on the interpretation of our results, comparative simulations were also carried out for linear and third order models (not shown here). 
While assuming only a linear inhomogeneity of the magnetization profile leads to a significantly worse fit of the static angle-resolved data, the fitted laser-induced dynamic change of the magnetization \textit{relative} to the equilibrium state describes the same trend within the experimental accuracy for both linear and third-order models. 
We thus conclude that a second order distribution is the simplest mathematical model to allow a consistent description of both static and time-resolved data, reflecting the asymmetric layer structure of our sample with distinct cap and seed layers, Ta and Pt, respectively.
While, strictly speaking, we can not rule out an inhomogeneity which is more complex than a second order modulation but leads to identical magnetic asymmetry spectra, this strongly suggests that the choice of the model does not impact our interpretation regarding an imhomogeneous demagnetization enhanced towards the interface with the Pt seed layer.

\begin{table}
	\fontsize{11}{22}\selectfont
	\centering
	\begin{tabular}{l@{\hspace{10\tabcolsep}}c@{\hspace{10\tabcolsep}}c}
		\toprule
		Layer	& $n$	& $\kappa$	\\ 
		\midrule
		Ta		& 5.1	& 4.5				\\
		$\text{Gd}_{24}\text{Fe}_{76}$ & 5.5 & 4.5		\\
		Pt		& 4.5	& 7.2				\\
		Glass 	& 1.5		& $<0.0002$			\\
		\bottomrule
	\end{tabular} 
	\caption{Real ($n$) and imaginary ($\kappa$) part of the complex refractive indices at 2.1\,\textmu m wavelength obtained for the individual layers and the substrate of the studied sample from MIR/NIR and NIR/VIS ellipsometry measurements. \label{tab:refractiveindices}}
\end{table}

As explained in the main article, this observation is in line with the depth-dependent absorption profile of the 2.1\,\textmu m pump pulse shown as an inset in Fig.~1(a) of the article, predicting the deposited energy to be largest within the Pt seed layer. 
The 2.1\,\textmu m differential absorption was calculated by a numerical simulation within the \textsc{udkm1Dsim} framework, taking multiple interlayer reflections into account that depend on the wavelength-dependent complex refractive index in the MIR to correctly calculate the amount of energy absorbed within each layer of the sample. 
The complex refractive indices ($\underline{n}=n+i\kappa$) of each layer at 2.1\,\textmu m wavelength were independently determined from MIR/NIR ellipsometry (home-built instrument \cite{Furchner_2020}) and NIR/VIS ellipsometry (Sentech SENresearch 4.0) measurements carried out under multiple incidence angles in reflection and transmission geometry on reference layer and substrate samples. 
The resulting values for $n$ and $\kappa$ that were obtained from analyzing the ellipsometry data and used for calculating the 2.1\,\textmu m absorption profile are shown in Suppl.~Table~\ref{tab:refractiveindices}.

\begin{figure*}
	\includegraphics[width=1\linewidth]{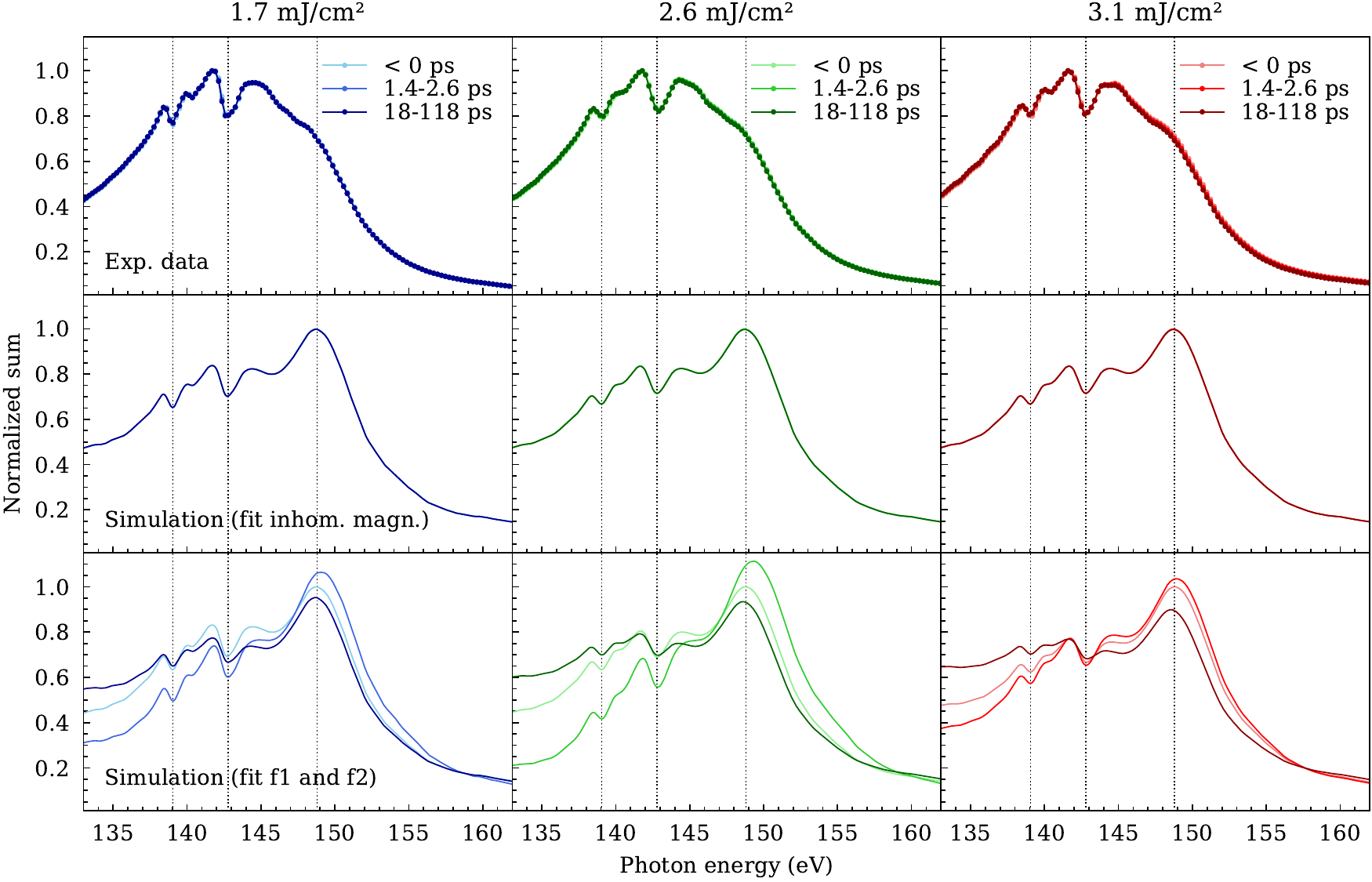}
	\caption{Time-resolved (non-magnetic) reflectivity spectra normalized to the maximum before 0\,ps as a function of excitation fluence and time after excitation (differently shaded blue, green and red curves, respectively).
	The experimental data (upper panels) and the simulations assuming an inhomogeneous demagnetization profile without any changes in the charge scattering factors $f_1$ and $f_2$ (mid panels) are nearly constant over delay time.
	In contrast, simulations assuming a homogeneous demagnetization with transient and element-specific changes in the charge scattering factors $f_1$ and $f_2$ (lower panels) show significant spectral weight shifts, which are not observed experimentally.
	Please note that the experimental reflectivity spectra are influenced by the envelope of the incident spectrum of the radiation source.
	While this leads to an intensity drop for photon energies $\geq$ 145\,eV with respect to the simulated spectra, it is not influencing the TMOKE \textit{asymmetry} which is normalized to the sum reflectivities.
	\label{Suppl_7_Figure}}
\end{figure*}

Finally, we want to show that it is not possible to explain the observed changes in the TMOKE spectra (peak shifts, photon energy-dependent response) that were attributed to an inhomogeneous demagnetization by only a transient change of the atomic/charge scattering factors, which would correspond to a purely electronic effect due to, e.g., a changed state occupation upon excitation. 
Please note that this is already highly unlikely because the free $4f$ states of Gd probed at the $\text{N}_{5,4}$ resonance are out of reach of the 2.1\,\textmu m excitation pulse and the other layers in the sample (i.e., Ta, Pt) possess no atomic resonances in close vicinity. 
In order to approximately model such effects, the demagnetization was assumed to be homogeneous, but the charge scattering factors $f_1$ and $f_2$ were allowed to be scaled individually for each element in the sample in order to fit the simulated magnetic asymmetry to the transient spectra (fits not shown). 
Compared to the purely magnetic \textit{homogeneous} simulations shown in Suppl.~Fig.~\ref{Suppl_6_Figure}, this allows to obtain a much better reproduction of the time-resolved asymmetry due to the larger number of free parameters.
However, it predicts significant changes in the sum reflectivity spectra (containing only electronic contributions), which are clearly not present in either the experimental data or the inhomogeneous demagnetization simulations (see Suppl.~Fig.~\ref{Suppl_7_Figure}). 
It can thus be ruled out that the observed effects are of non-magnetic origin, which confirms our conclusions regarding an inhomogeneous transient magnetization profile.

\renewcommand*{\bibfont}{\fontsize{11}{22}\selectfont}

\end{document}